# Optimizing achromaticity in metalenses, and development of a layered thin-film metalens

Calvin M. Hooper [a], Sarah E. Bohndiek[a,b], Calum Williams[a,*]

[a]Department of Physics, Cavendish Laboratory, University of Cambridge, JJ Thomson Avenue, Cambridge, CB3 0HE, UK;
[b]Cancer Research UK Cambridge Institute, University of Cambridge, Robinson Way, Cambridge, CB2 0RE, UK;  *cw507@cam.ac.uk

## ABSTRACT

Metalenses are ultrathin optical devices designed to replicate behavior of conventional refractive lenses, or lens arrays, utilizing nanoscale resonant structures to redirect incident light. These are often comprised of discrete meta-atoms such as nanoscale dielectric pillars. Achromatic focusing—associated with traditional multi-element refractive counterparts—is frequently attempted with single-layer metalens designs, which has proven difficult to achieve with bounded refractive indices and total lens thickness. A recent study (F.Presutti and F.Monticone, 2020) formalized this, applying optical delay-line limitations to metalenses, resulting in a generalized trade-off in achromaticity for focusing systems.

In this work, we (1) theoretically explore achromaticity in metalens design, and (2) propose a thin-film multilayer design as an alternative to the discrete meta-atom approach for large numerical aperture (NA) achromatic metalenses. It is shown that wavefront modulation can also be achieved with spectrally-varying transmission magnitudes, rather than purely matching a phase profile. In fact, even with a bounded refractive index, perfect achromatic operation over a given spectral range can be offset by imperfect operation elsewhere, resulting in a NA limited by the smallest general spectral feature controlled. These considerations lead to a generalized phase-matching optimization routine, and a thin-film metalens is simulated, utilizing layered TiO2/MgF2 with total thicknesses ≤1 μm (≤20 layers), focusing across 6 simultaneous wavelengths (350-740 nm, Δλ~65 nm). A significant proportion (>40% spectral average) of the reflected light is focused for moderate NA (~0.35). With the maturity of the optical coating industry, the conformal thin-film approach reduces manufacturing complexity from its discrete nanoscale meta-atom equivalents.

**Keywords:** achromatic, bandwidth limit, quasi-achromatic, subwavelength thin-films, multispectral, metalens, multilayer films, optimization.

## 1. INTRODUCTION

Compact lenses are desirable in a variety of miniaturized optical imaging and sensing applications. From industrial inspection to minimally invasive endoscopy, spatial limitations restrict the complexity of possible lens elements and imaging lens assemblies[1]. For such applications, either a reduction in imaging performance (i.e. increased chromatic aberrations) is tolerated or a maximum physical size limitation is placed on the system in order to maintain a certain level of performance. An ultrathin lens should mimic functionality of a conventional lens / assembly (objective) but in a highly reduced form factor[2]. One such important functionality is achromaticity; in refractive optics, achromaticity is conventionally achieved through multiple cemented lens elements (i.e. doublets, triplets) using several materials, to mitigate chromatic aberrations. Replicating this capability in ultrathin lenses is challenging.

Within the field of ultrathin lenses, a divide is often drawn between diffractive lenses and the increasingly popular *metalenses*. The distinction in the literature between the two is often somewhat arbitrary—and may even be unhelpful in certain circumstances—but a reasonable distinction might be that metalenses explicitly make use of resonant structures to control the propagation of light, whilst diffractive lenses do not[3]. This is still very much an oversimplification (almost all diffractives will exhibit *some* degree of resonance), but under this definition the ultrathin lenses considered here would be classed as metalenses.

## 1.1 Achromaticity limit

A fundamental motivation of ultrathin metalenses is to replicate and augment the capabilities of conventional *thick* refractive lenses or lens assemblies, such as achromatic focusing and achieving large numerical apertures (NA). Achieving achromaticity has, however, proven to be difficult to realize in high-efficiency structures with a large NA. In the case of perfect lensing—where normally incident light is focused through a perfectly transmitting medium—it can be shown that the spectral nature of an applied phase shift must be such that the lens exerts a constant group delay on the light[4,5]. Equivalently, this can be viewed as the metalens causing the wavefront to experience some spatially varying optical path difference (OPD) across its surface.

Once the problem of designing such a metalens has been identified as being equivalent to applying an OPD, a challenge arises as to how this can be physically achieved without requiring arbitrarily high refractive indices or large thicknesses. This turns out to be difficult to achieve in an achromatic fashion, something which can be seen particularly easily for a single resonator: a high group-delay implies strong confinement, implying low radiative losses, resulting in a high quality factor, and strong chromatic aberrations. This notion has been formalized in a recent paper[5] by drawing analogy to optical delay lines (see Figure 1), but it seems empirically likely that this holds for a far broader set of conditions. As such, a section of this work focuses on achieving achromatic metalenses, with arbitrarily large NAs, but without reliance upon arbitrarily high refractive indices or arbitrarily large physical lengths.

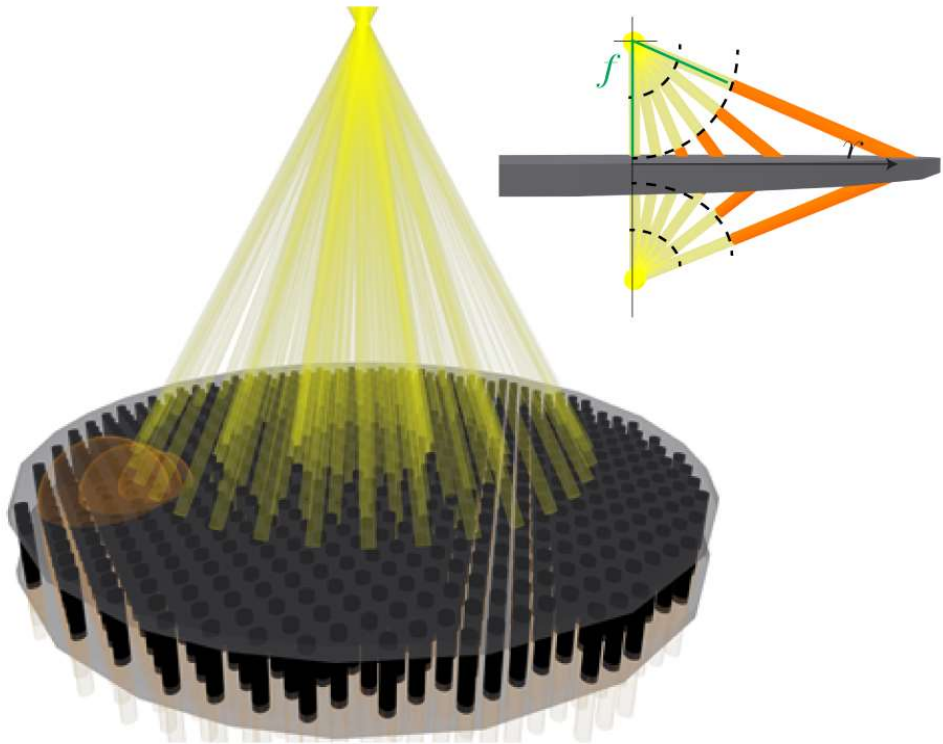

Figure 1: A schematic diagram of a metalens as a series of optical delay-lines. The surface of the lens is shown by the grey layers, with the delay-line structures between the two. A Huygens construction in 3D (left) is shown in orange, demonstrating how tuning the delay lines to separate delays results in a bending of the output light to a focus, as shown. This is highlighted in the small inset (right).

## 1.2 Meta-atoms and metalens design processes

In order to produce any such optical modes, nanostructured resonant cavities known as meta-atoms are used. In combination, these meta-atoms form metasurfaces, which in this instance takes the form of a metalens (Figure 2A). To control light independently across a surface, these resonant cavities are often designed to decouple the resonance of a particular meta-atom from that of its surroundings, typically resulting in a series of discrete meta-atoms positioned across a surface[3]. In general, however, and especially when a metasurface with some continuous property is desired, this is not a necessary feature of metalens design, and more continuous structures may provide some benefits. As there typically exists a tradeoff between the volume of the structure controlled to produce a mode, and its quality factor, one might even expect a potential improvement in the spectral control available to extended structures. Indeed, in highly optimized metalens designs without practicality constraints, increasingly continuous structures have been observed[6].

Typically, two sets of processes are used in metalens design, known as forward- and inverse- design, which are summarized in Figure 2.

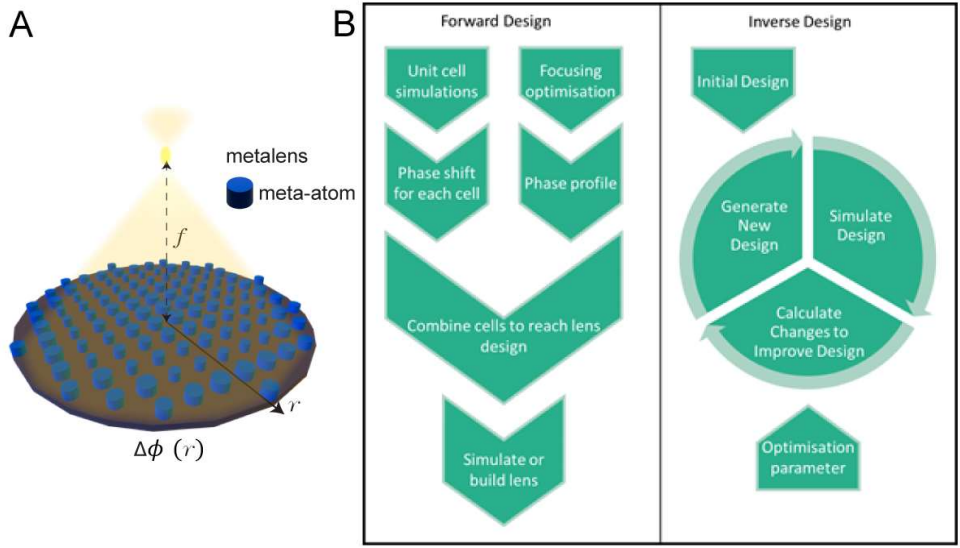

Figure 2: A- Schematic representation of a metalens: discrete meta-atoms with varying properties are distributed across a surface, resulting in a spatial variation of the phase of light propagating through the material. B- Metalens design paradigms: Forward design splits (potentially approximately) the problem of lens design into two parts: designing the building blocks, or meta-atoms, of a metalens, with a phase profile designed in parallel to optimize some feature of the focusing (such as, for instance, numerical aperture at high fields of view). Inverse design is a more holistic process, whereby the operation of the lens as a whole is considered. This is more time consuming, with typically less intuitive results, and less ability to prove optimality, but is often the only option to avoid approximations.

Forward design relies on splitting the problem of metalens design into more manageable, and approximately separable, subproblems. Typically, this involves designing an idealized phase profile, then simulating a library of meta-atoms to attempt to approximate the properties of this phase profile. The benefits of this are: the problems of phase profile and meta-atom design become relatively straightforward in many cases; the design process is fast; the result is typically intuitive; and one where guarantees of global optimality may be possible. A disadvantage, however, is that metalens design is rarely truly separable: meta-atoms interact with their neighbors to some degree, and which meta-atoms are their neighbors is a function of the choice of phase profile. The increasingly common selection of waveguide-like meta-atoms helps to decouple resonant modes from that of the atom's neighbors, but does not completely do so[3]. Forward design also often requires the optimization processes of inverse design—particularly in the selection of meta-atoms with a large number of degrees of freedom, where completely sweeping the space of possible meta-atoms becomes impractical. The primary objective here was the forward design of a metalens of a given phase profile. Whilst custom wide-angle phase profiles were considered, and may be an interesting focus for future work, the standard hyperbolic phase profile was selected:

$$\Delta\phi = -\tfrac{2\pi\Delta L}{\lambda} = \tfrac{2\pi}{\lambda}\left(f - \sqrt{f^2 + r^2}\right) \quad (1)$$

where $\Delta\phi$ is the phase difference induced by the metalens at a point $r$ for a wavelength of light $\lambda$ to be focused to a focal point at $r = 0$ and a distance $f$ from the plane of the lens. This can also be simply derived by considering the extra path difference $\Delta L$ between a point on the lens and the focal point. It can be thought of as the phase profile which solely optimizes NA at normal incidence.

## 1.3 This work

Methods of focusing for large OPDs, even with constraints on refractive indices and lens thickness are considered. It is demonstrated that the primary limiting factor in metalens design is that of the gradient of spectral features which can be engineered into a material, and that while large path differences and high refractive index contrast are common, they are not the only methods of realizing these features. Additionally, it is shown that focusing to multiple points, even for a constant focal plane, can reduce the spectral gradients needed. It is noted that these solutions are typically lost in standard forward design processes whereby meta-atoms are chosen to match the phase of an ideal phase profile, and, instead, a second approach is proposed: optimizing for a more general correlation between the ideal phase profile and the achieved transmission coefficient. This is demonstrated in a heavy approximation scheme for a multilayer meta-atom composed of alternating high-low index thin-films, and is shown to provide simultaneous focusing across 6 UV-VIS-NIR wavelengths, as well as resulting in an interesting optimized multilayered structure seeming to approximate continuous layers, the potential of which is discussed further.

## 2. ACHROMATIC FOCUSING WITH BOUNDED THICKNESS AND REFRACTIVE INDEX

As noted in 1.1, focusing may often be considered as realizing a particular refractive index or *effective* refractive index profile to realize some target transmission profile. In the case of achromatic focusing, the transmission profile can often be more simply considered to be an achromatic OPD profile, rather than a general phase profile. Implementing arbitrary achromatic path differences has, however, proven to be difficult in bounded refractive index materials. In this section, two methods of achieving achromaticity are proposed. Both methods are not without their limitations: in particular, the performance of each method is limited by the spectral control which can be exerted by a meta-atom.

### 2.1 Multiple focal points

Whilst the notion that the *bending* of light to a single focus requires an OPD is powerful, there is a large class of lenses which can be considered for which this is not their mode of operation. It has already been noted[4] that this gives rise to the inapplicability of the limit[5] for conventional diffractive lenses. Considering the example of an amplitude-only Fresnel zone plate (FZP), the mechanism involved in focusing is not to modulate a wavefront into meeting a particular criterion (i.e. having the correct phase to focus), but rather to simply remove sections of the wavefront which don't meet that criterion sufficiently closely. This behavior typically this comes at the expense of a reduction in focusing efficiency. An alternate standpoint from which to consider FZPs is that they arise from attempting to optimize a metalens for a given phase profile (typically the hyperbolic phase profile considered above) but with a library of meta-atoms limited to a perfect transmitter and a perfect blocker of light (whether reflective or absorptive). A direct approach merely trying to find a meta-atom applying the correct phase shift for every particular point on the lens would have significant difficulty here, given that the library of meta-atoms does not contain a complete range of phase shifts. A more general optimization routine can be considered, however, optimizing for focusing efficiency even when the meta-atoms used exhibit imperfect transmission, and would return something resembling a FZP. Whilst the FZP is fundamentally chromatic in its design, its underlying principle—achieving focusing without relying on phase shifting—is a notion which can be brought forward more generally.

Adding imperfect transmission to this design had a secondary effect: whilst the phase profile which was being optimized was designed to focus light to a single point, the optimal solution for the particular meta-atom library provided, in fact contained an infinite number of diffracted focal points. As a result of this, whilst the FZP does fail to provide a wavelength-independent focal length, it is no longer bound by the achromaticity limit proven earlier, as its mode of operation is not that of applying a single path difference to a beam of light. In fact, this is all equivalent to the statement that this limit only applies for this particular phase profile (or perhaps a piecewise combination of phase profiles such as this one but with different target focal lengths or translations), since any particular lens can be viewed as focusing an input wave to a series of points of varying phase and intensity. And, as such, the notion of adding additional focal points is a powerful one, as it encompasses not only the Fresnel-like diffraction patterns (i.e. an infinite series of focal points), as has been previously considered[4], but can also be used to describe any other phase profile optimized for any other purpose, or, more generally, any transmission profile. A simple illustrative example—building on the strength of the FZP's multiple diffraction orders, whilst incorporating the smooth spatial control of properties typically associated with metalenses—can be designed with the following transmission profile, arising from the superposition of multiple phase profiles.

$$t(x,\lambda) = \frac{2+e^{i\frac{2\pi}{\lambda}\left(f-\sqrt{f^2+x^2}\right)}+e^{-i\frac{2\pi}{\lambda}\left(f-\sqrt{f^2+x^2}\right)}}{4} = \cos^2\left(\frac{\pi}{\lambda}\left(f-\sqrt{f^2+x^2}\right)\right). \tag{4}$$

Here the first term corresponds to a beam passing straight through, the second to a beam diverging as though normally incident light were arriving from the *wrong* side of the lens, and the final term corresponds to focusing, where all these terms were selected to ensure the contribution of the final term was maximized, whilst the total sum remained real and between 0 and 1. This profile exerts no phase shift, and hence can't be limited by any achromaticity limit based on OPD. It does, however, require that arbitrarily large spectral refractive index gradients be engineered at arbitrarily large distances from the focal point. Of course, varying transmission coefficients without varying phase, at least in the context of thin-film optics, is practically difficult (although polarization sensitive meta-atoms between crossed-polars could have an equivalent effect). In practice, constraints on transmission from meta-atom libraries are never this simple, but this still acts to demonstrate how, even with a limited meta-atom library, unable to exert an optical path difference, it is possible to strongly focus light

A second, more interesting, example is also considered, demonstrating how a change in the spot-shape can avoid achromatic OPDs: a 1-dimensional lens focusing light to two focal points a distance $\delta$ apart, and precisely $\pi$ out of phase. It should be noted that the authors of the aforementioned achromaticity limit[5] did briefly note that the limit would fail to apply for varying spot shapes such as the one we consider, here. The action of this lens is no longer the simple phase profile initially presented, and in fact can no longer even be described by a single phase profile, it is instead described by a transmission profile:

$$t(x,\lambda) = e^{\frac{2\pi i}{\lambda}\left(\sqrt{f^2+(x-\frac{\delta}{2})^2}\right)} - e^{\frac{2\pi i}{\lambda}\left(\sqrt{f^2+(x+\frac{\delta}{2})^2}\right)}, \tag{5}$$

where $x$ is the point on the lens for which we're considering the transmission. This demonstrates how a metalens, not mimicking the behavior of a FZP, may provide wavefront modulation other than a pure OPD to produce a diffraction-limited image, albeit one with a somewhat increased spot size. Indeed, if we allow $\delta$ to vary with $\lambda$ (noting that, whilst the focal points move to maintain a spot size roughly double that of the diffraction limit), we are able to let $\delta = \lambda$, which results in a lens where transmission tends to zero as the distance from the focal point increases. Expanding for large $x$, we find:

$$t(x,\lambda) \propto e^{\frac{2\pi i}{\lambda}\left(x+\frac{f^2}{x}+O(x^{-2})\right)} O(x^{-2}). \tag{6}$$

Considering this expression, we can see that, at large distances, something akin to a path difference will be required, likely still obeying a similar limit to that discussed above. However, from the decreasing contribution of these higher distances, it is also clear that more freedom can be expected in realizing them.

The generalization of phase profiles to transmission profiles also raises an interesting observation and line of reasoning. Typically, when designing metalenses, a library of meta-atoms exerting varying phase shifts is produced, maximizing transmission for the meta-atom chosen to realize each phase shift. However, whilst the phase shift described by (1) remains the ideal case, there is no guarantee that maximal focusing efficiency is obtained by individually maximizing transmission for the correct phase shift at each position. For an extreme example, consider the case where the meta-atom library only contains meta-atoms realizing a single phase, but otherwise arbitrary transmission coefficients as in (4). If the aim of optimization is simply to realize the phase profile with maximal transmission at every point, then all possible metalenses designed would be regarded as equal. However, in practice, significant focusing efficiency could still be achieved with the following phase profile:

$$t(x,\lambda) = \frac{2+e^{i\Delta\phi(x,\lambda)}+e^{-i\Delta\phi(x,\lambda)}}{4}. \tag{7}$$

Whilst the authors know of no general limit on transmission coefficients similar to the delay line limit on OPD, it should be expected that some limitation, in a similar vein to the group-delay limit, should exist. We might expect, however, it to be difficult, in general, to maintain large $\frac{dt}{d\lambda^{-1}}$ over a wide range of $\lambda$, meaning that a general reduction in transmission coefficient ($t(\lambda) \mapsto \alpha t(\lambda) \forall |\alpha| \leq 1$) might be expected to lie more easily within such a limit. We suggest this, because it notes a second interesting property of (6), which is that general approximations to it, for a meta-atom library without phase shifting ability (as described in 7) is potentially made significantly easier for large $x$ due to the decrease in the transmitted amplitude, there. In fact, an analogous equation to (4) could be produced:

$$t(x,\lambda) = \frac{2+A\cos\left(\frac{2\pi}{\lambda}\left(x+\frac{f^2}{x}+O(x^{-2})\right)+\phi\right)O(x^{-2})}{4}. \tag{8}$$

for some $A$ and $\phi$ optimized for a passive metalens.

In summary, the addition of extra focal points, even simply to vary the shape of the resulting focal spot, can provide a huge additional degree of freedom, allowing this particular achromaticity limit to be effectively bypassed, and requiring a more general limitation on transmission coefficients. In addition, this is not merely confined to adding points to varying focal planes at increasing distances, as has previously been considered[4]. This set of lenses might be classed as either diffractives or metalenses, depending on the definition used[3,4], suggesting that perhaps such a stark divide is unhelpful.

**2.2 Spectral dead-zone**

Whilst the previous method adds a degree of freedom into metalens design by utilizing imperfect transmission, this method considers how, in general, even with a bounded refractive index and thickness, arbitrarily large path differences may be realized under the condition of arbitrarily sharply designed spectral features, and, where this isn't possible, the limitations which may apply. A common theme between both this and the previous section is that metalenses lack any implicit requirement as to the shape and localization of the pulse arriving; a feature not shared by delay-lines. Perfectly realizing the phase profile for a single focal point, it may seem that pulse-shape should be maintained, but in any system with imperfect transmission, as has been shown, the resulting optimized phase profile may not share this property, and hence may not be bound by such a limit.

In this section, we consider how, for any particular point on the lens, precise engineering of a refractive index dispersion curve may allow for achromatic performance for arbitrary path differences, aside from a discrete number of achromatic 'spectral dead-zones' where a highly chromatic and highly *incorrect* path difference (relative to an idea path difference profile) over a small spectral range allows for achromatic performance elsewhere. Consider a medium with refractive index $n(k)$ over which complete control is available, and consider a length $l$ within it. The difference in phase between a beam passing through this medium and one passing through the vacuum is given by $\Delta\varphi = (n(k)-1)kl$. The OPD $\Delta\mathcal{L}$ between the two has the property that $\Delta\varphi = k\Delta\mathcal{L}$. Hence, if a particular achromatic OPD is desired, $n(k)$ can be defined as:

$$n(k) = 1 + \frac{\Delta\mathcal{L}}{l}. \tag{9}$$

Alone, this borders on the trivial, suggesting that, as expected, to achieve a particular achromatic path difference, the product $n(k)l$ should need to be arbitrarily large in order to produce arbitrarily large $\Delta\mathcal{L}$. A subtlety, however, is added by considering the fact that the phase difference is not physically unique, and a more general phase difference can be given by $\Delta\varphi = (n(k)-1)kl + 2m\pi$ for some integer $m$. Converting this to a path difference, and calculating the required refractive index to achieve this, we obtain:

$$n(k) = 1 + \frac{\Delta\mathcal{L}}{l} + \frac{2m\pi}{kl}. \tag{10}$$

Where the integer $m$ may be chosen freely and independently for each wavelength. One particular choice, which retrieves constant spacing between spectral dead-zones is given by setting $m$ such that $n(k)$ is larger than unity by the minimum amount for any given wavenumber, or equivalently:

$$n(k) = 1 + \mathrm{mod}(\tfrac{\Delta\mathcal{L}}{l}, \tfrac{2m\pi}{kl}). \tag{11}$$

Here, the OPD applied to the wavefront is well-defined and equal to its intended value almost everywhere, with the exception being the points where $\mathrm{mod}(\frac{\Delta\mathcal{L}}{l}, \frac{2m\pi}{kl})$ is discontinuous, and the gradient is undefined. At these points, the phase shift applied is not meaningfully controlled or defined. In practice this is not realizable, both due practical difficulty and the fact that the Kramers-Kronig relations have been neglected. However, when considering the application of a general wave equation as has been used previously in deriving achromaticity limits[7], this is still a valid choice.

For low $k$, this design provides no benefit, but, as this is bounded by $n(k) \leq 1 + \frac{2\pi}{kl}$, for any combination of length and maximum refractive index, a $k$ can be found beyond which this medium operates successfully with a limited length and bounded index.

Whilst neglecting the Kramers-Kronig constraint may initially seem like it would significantly reduce the degrees of freedom of the problem, a constraint on the imaginary part of the refractive index was already in place from the assumption that the refractive index was real, and this merely exchanges that constraint for one which is more complicated to calculate, and an optimization problem is less obvious than simply matching to a phase, with the magnitude of transmission coefficients now varying. However, as demonstrated in 2.1, it is entirely feasible to achieve focusing in this way. Additionally, these constraints will primarily cause attenuation localized near the spectral discontinuities – where the behavior of light was already worst defined.

Both this regime and that presented in 2.1 find the length scale required for the OPD in the gradient with respect to wavenumber of a dimensionless quantity, whether transmission coefficient or refractive index, rather than in the scale of the medium itself. As a result of this, properties which have a tendency to vary rapidly with wavenumber are seen. Whilst the ability to engineer spectral features with an incredibly low bandwidth has been demonstrated[8-10], even within subwavelength volumes, the authors are unaware of any work demonstrating complete control over a transmission or refractive index spectrum over a range of frequencies without increasing the structure volume (e.g. multi-layer thin films, effective index media etc.). This inability to design arbitrary spectral features within a finite volume can be characterized by defining $\Delta k$ as the minimum spectral resolution we're capable of engineering into a spectrum. In the specific example considered above, spectral dead-zones, spaced by $\frac{2\pi}{\Delta \mathcal{L}}$ cannot be designed with a size less than $\Delta k$, resulting in a non-negligible range of wavelengths which are incorrectly focused. As $\Delta \mathcal{L}$ increases, eventually the spacing becomes less than $2\Delta k$, meaning that the correct gradient and the correct spectral dead-zone can no longer both be realized. In practice, this means the focusing efficiency contributed from path differences around $\Delta \mathcal{L}$ reduces, being bounded by $1 - \frac{\Delta k \Delta \mathcal{L}}{2\pi}$. The fact that performance is bounded by the minimum spectral feature size which can be engineered seems to be a likely generalization to achromaticity limits based solely on path difference, and represents a region for future work. It also suggests the power of utilizing coupling to other systems with differing length scales and dispersion relations[11,12].

## 2.3 Optimization

Over the course of this research, the authors were unable to find examples of scattering properties being controlled with complete freedom over a large bandwidth, without the usage of large cavities, suggesting that some empirical limit underlies this (or possible even a more fundamental limit – a topic for further investigation), meaning that perfectly designing a meta-atom to match a completely general scattering profile becomes a problematic task. It has been further shown that even purely matching transmitted phase to a phase profile will not, in general, produce optimal results (as was discussed earlier for a FZP). As a result, to take full advantage of the degrees of freedom still left within our library of meta-atoms after accepting some fundamental limitations, a more nuanced approach is desired. In particular, if a phase of $\Delta \varphi$ is desired at a particular position and wavelength, whilst a set of meta-atoms, each with transmission coefficient $t$ are provided, how, in general, can the best transmission coefficient be chosen? Roughly speaking, it can be assumed that any contribution to the complex amplitude in quadrature with $\Delta \varphi$ does not contribute to focusing, and hence $t$ can be chosen: Therefore, at any point on the transmission spectrum, we can find $t$ to maximize

$$\Re\{te^{-i\Delta\varphi}\}. \tag{12}$$

Applying (12) to calculate a single optimized quantity across the entire lens, we can simply integrate across the lens' surface, and find a structure providing a transmission function $t$ which optimizes $F_\lambda$ defined as follows:

$$F_\lambda = \int \Re\{t(x,y,\lambda)e^{-i\Delta\varphi(x,y,\lambda)}\}dA, \tag{13}$$

where $F_\lambda$ is the function of optimization for a particular wavelength, and $x, y$ denote the position of an area element $dA$. This still applies only at a single wavelength. The most accurate way to create an optimization coefficient across a range of wavelengths, from this, would be to optimize:

$$F = \int |F_\lambda|^2 d\lambda. \tag{14}$$

However, since meta-atoms are typically localized in position space, rather than frequency space (photonic crystals achieve the latter in some sense, but are necessarily required to be thick to achieve their properties, and so are not of interest, here), it would be advantageous to express the function of optimization as:

$$F = \int F_A dA. \tag{15}$$

This can be achieved by deciding to instead to find a structure, and hence $t$, to maximize a lower bound $F_{\text{low}} \leq F$, where:

$$F_{\text{low}} = (\int \int \mathfrak{R}\{t(x,y,\lambda)e^{-i\Delta\varphi(x,y,\lambda)}\}dA)^2. \tag{16}$$

The following can then be maximized:

$$F' = \int \int \mathfrak{R}\{t(x,y,\lambda)e^{-i\Delta\varphi(x,y,\lambda)}\}dA\,d\lambda. \tag{17}$$

The decision to maximize the positive root of $F_{\text{low}}$ arises from the fact that the target is to be in phase with the provided profile. This is clearly a significant approximation, but allows the order of the integrals to be exchanged. Since the transmission coefficient of a meta-atom at one position has no bearing on that at another position (at least in the typical forward design approximation regime); the following spatially separated meta-atoms can all be optimized independently, with the resulting meta-atom at any particular position $(x,y)$ chosen to be:

$$\operatorname*{argmax}_{\alpha \in \text{meta-atoms}} \left( \int \mathfrak{R}\{t(\alpha,\lambda)e^{-i\Delta\varphi(x,y,\lambda)}\}d\lambda \right). \tag{18}$$

This form is heavily approximated, and should not be expected to perform particularly well. In addition to the approximations considered above, this only achieves the best possible matching to an optimized phase profile, but it can't generally be assumed that maximizing matching to an idealized profile actually maximizes focusing efficiency when the meta-atom library is insufficient to allow for an ideal matching. It also allows the optimization algorithm perfect freedom to trade off performance at one frequency for another, something which is often not desirable. Fortunately, this can be significantly improved by simply raising the contents of the integral in (18) to a real power greater than unity.

Potential improvements can be made, centering around finding the reciprocal basis set of transmission coefficients in the nearfield corresponding to a complete set of focal points in the farfield. Unfortunately, this is difficult to achieve analytically, and is closely related to the problem of deconvolution, making it incredibly sensitive to rounding errors during computation. An improvement upon this method, whilst still within the domain of a forward design process, would be to rigorously simulate the propagation of light away from the boundary, and measure the intensity within a target pixel. Depending on the constraints of the problem, if this pixel were large enough, this would allow designs such as that proposed in (5) to arise naturally. Since this propagation would occur in a vacuum, the analysis would be relatively straightforward. One feature still neglected, however, is the interplay between adjacent meta-atoms. This could only be completely assessed through complete inverse design, although there is the potential for approximations to a gradient in meta-atom properties by measuring the effect of adjacent meta-atoms.

## 3. CONFORMAL MULTILAYER MULTISPECTRAL METALENS

A common feature of metalens design is the notion of discretization—often an attempt is made to ensure that all meta-atoms resonate as independently from their neighbors as possible. Dielectric meta-atoms, however, must be closely packed to remain beneath the wavelength of light (at least for common dielectric materials) and given the close packing between such meta-atoms, the evanescent fields from each meta-atom cannot have time to fully decay before reaching the adjacent meta-atom, resulting in a necessary degree of coupling[3].

In this section, the notion of a conformal multilayer metalens (Figure 3) is presented as it arose—as an assumption and then result of designing thin-film cavities to test the optimization routine presented in the previous section. It will be seen however that, due to the continuous properties of the lenses, a conformally layered design appears to arise naturally, and presents potentially useful properties. Whilst the approach and approximations used here were reasonably crude, reasonable focusing was demonstrated across multiple wavelengths (through UV-VIS-NIR) making this forward design process an incredibly efficient method to calculate the starting point for less approximate inverse-design methods. Without any techniques available to independently control multiple small-bandwidth spectral features of the transmission curve at a subwavelength scale, any attempts at complete achromaticity using the above methods would be infeasible. Instead, the optimization process arrived at is tested for a metalens with multispectral (rather than broadband achromatic) imaging capabilities. For optical imaging applications, however, this tradeoff can be considered tolerable: as firstly, a complete hypercube's worth of data (i.e. 2D spatial dimensions + 1D of continuous spectral information cannot yet be recorded on a single 2D pixelated detector; and secondly, it is common to deposit thin-film spectral filters atop detectors to selectively discriminate the light (i.e. the ubiquitous red, green, blue Bayer filter, or multispectral filter array) and hence a metalens which is only targeted to these filter functions may be sufficient.

### 3.1 Thin-film meta-atom design

To obtain sufficient degrees of freedom to simultaneously modulate 6 separate wavelengths, thin-film cavities were used as meta-atoms in a metalens design. The materials used in the thin-film cavity design were the widely utilized low-high index dielectrics $MgF_2$ and $TiO_2$, with values for their dispersive refractive indices and extinction coefficients used between 200-1000 nm[13,14]. We note that for manufacturing simplicity, these two dielectrics can be replaced with two polymers with low and high refractive indices, and molded in a similar way to cost-effective polymer lenses (see 3.3).

Thin-film structures provide certain benefits and disadvantages to the design process. Two particular advantages brought are conceptual simplicity, and ease of simulation, with existing functions already having been implemented[15], both of which lend themselves towards a fast implementation. One drawback, however, is the strong coupling between adjacent meta-atoms, which could not be accounted for this separable forward design process but can be incorporated in future work. Since the optimization function varied continuously, it was assumed that adjacent structures would be continuous (an assumption which was borne out by calculations), and hence could be taken to be approximately flat over sufficiently short length scales. However, bleeding between adjacent meta-atoms was expected to be significant.

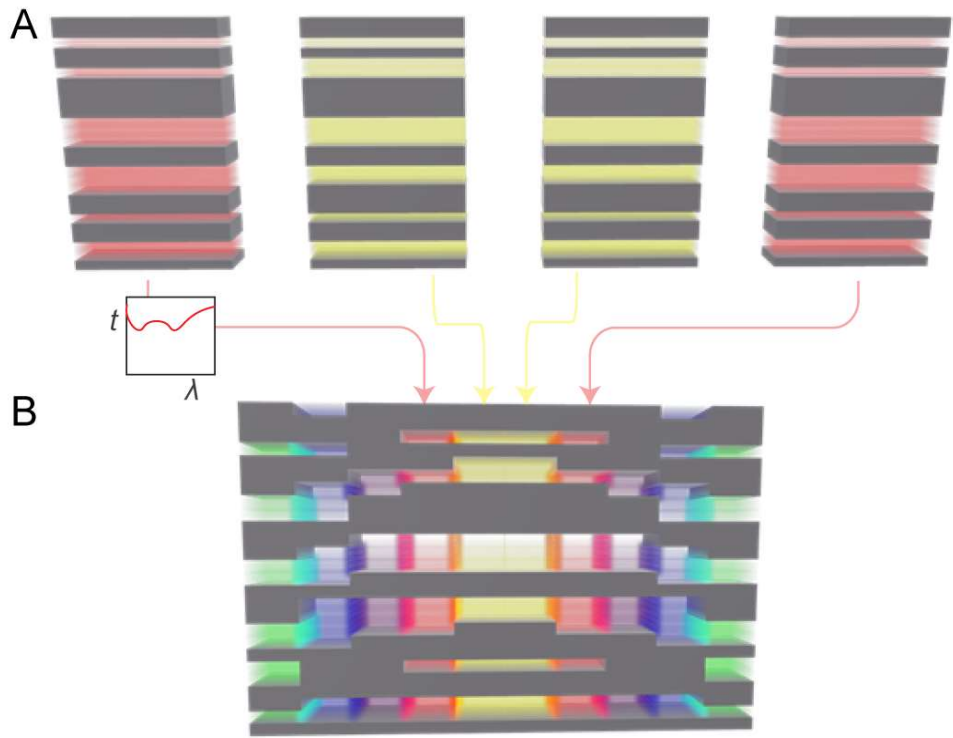

Figure 3: Multilayer metalens concept using thin films whereby each meta-atom here is a thin-film stack with specific transmission function, which in combination produce multispectral focusing. A- demonstrates the approximation used in forward design: a series of large meta-atoms are produced independently from one another, each assuming itself to be infinite in extent. B- demonstrates how, in practice, coupling and mixing occurs between two adjacent meta-atoms, significantly disrupting the resonant properties from those intended. An extended version of this structure would constitute a multilayer metalens design.

With some minor optimizations added to calculate thin-film reflections more efficiently in for this particular use-case, MATLAB's built-in `fmincon` minimization routine was used to find a thin-film structure with a transmission coefficient maximally matching (as described in section 2.3) the standard hyperbolic phase profile of (1). Whilst this optimization method was only local, and hence relied heavily upon a well-chosen initial condition, it was found over multiple global searches that the best value obtained was nearly always that locally optimized from initially evenly spaced layers.

## 3.2 Optimization Results

The aforementioned optimization routine was utilized for two distinct reflective metalenses, for two slightly varied sets of wavelengths, and two different numerical apertures (although this was unable to affect the separable meta-atom optimization routine). This produced a refractive index profile which could then be exported to a FDTD solver[16]. An example cross-section of the high-low index structure is shown in Figure 4A. Additional optimization routines were also carried out to assess the performance of the lens as a function of an increasing number of thin-film layers, for a particular set of wavelengths (across UV-VIS-NIR; 350-740 nm), and an optical path length chosen to be much larger than any of them. These results showed a general trend of improvement at a decreasing rate, eventually reaching a constant value, and indicates after approximately 20-30 layers, there is generally negligible improvement in performance. On top of this general trend was also some additional variation due to the different local minima found by the optimization routine. This is shown for one such set of wavelengths in Figure 4B. In fact, for almost all combinations of wavelengths and almost all large OPDs, the precise choice of path difference matters little, since the phase shifts required of each wavelength will typically be mutually unrelated.

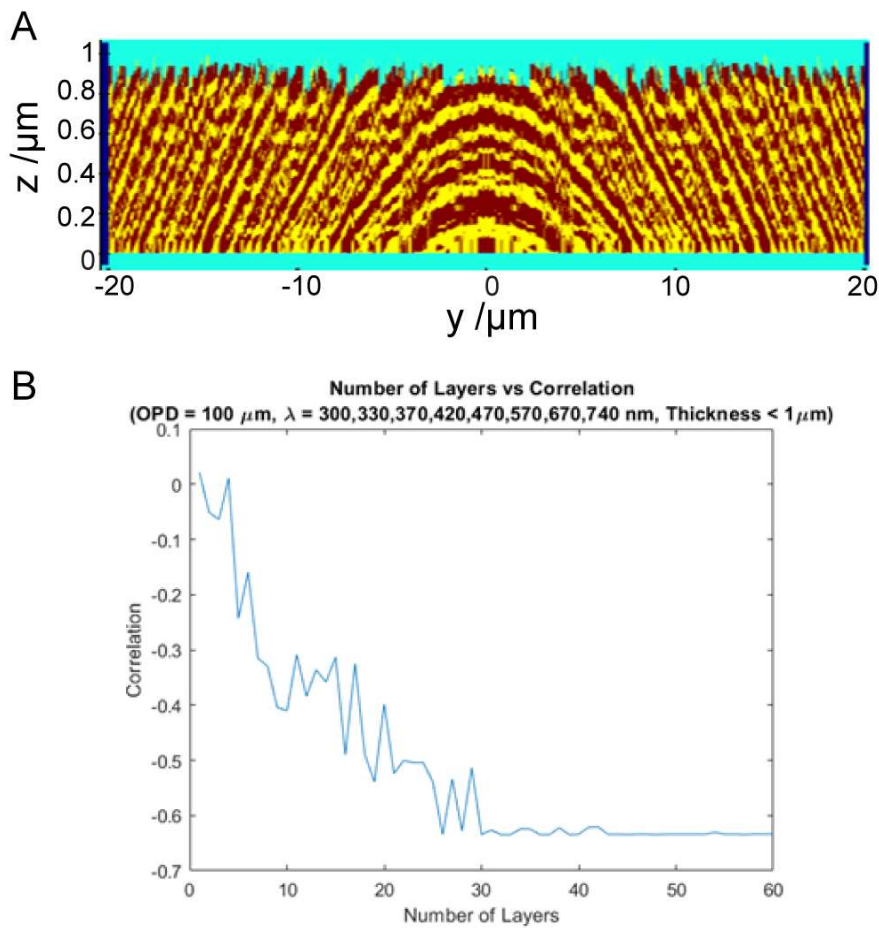

Figure 4: Optimized metalens design. A- An example refractive index profile of the design over a significantly larger range than was considered later (due to the increasing computation times with lens size). The yellow, dark red, and cyan colors refer to $MgF_2$, $TiO_2$ and air, respectively. Note the formation of generally relatively continuous layers of increasing slope further from the center. However, with shorter wavelengths in particular, this slope would result in a potentially significant horizontal translation due to refraction across the surface of the lens, and coupling between adjacent meta-atoms. It should also be noted that the gradient of the apparent bands of dielectrics is accentuated by the scale, and does not actually exceed unity within this region. B- The improvement in the optimized correlation coefficient (the negative of eq.18) as the number of layers provided is increased for a set of 8 wavelengths. The value of the correlation has been defined to be negative simply to allow direct substitution into MATLAB's built-in optimization toolbox, which is designed for minimization problems.

These designs were then simulated in the nearfield with a FDTD solver[16] centered in a region of air, with PML boundary conditions and normally incident broadband light, and local resultant fields extrapolated to the farfield with a built-in function. The resulting intensity and focusing efficiency in the focal plane at 6 different wavelengths (350, 400, 470, 570, 670, 740 nm) is shown in Figure 5, with Figure 6 showing the performance with a larger metalens but with one design wavelength adjusted by +10 nm (410 nm). A diffraction-limited spot is overlaid for reference in each intensity plot. Both metalenses used only two different materials, and the meta-atoms were comprised of ≤20 layers, less than 1 μm in height, and with a square unit cell (base) of 150 nm x 150 nm. The focusing efficiency (not equal to the complete focusing efficiency discussed earlier) is measured by the power within the diffraction-limited spot relative to the reflected power within the total extent of the focal plane.

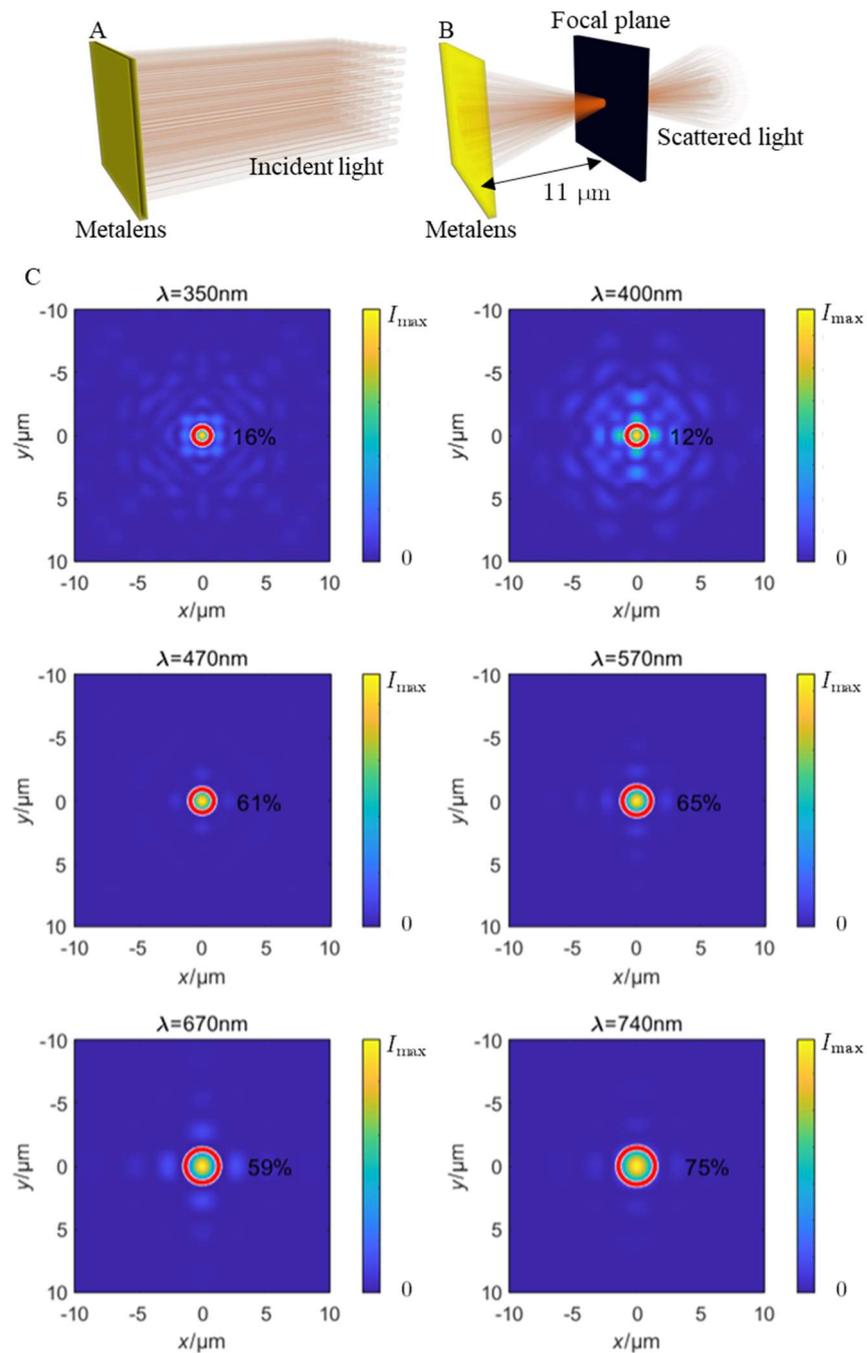

Figure 5. A,B- Light incident on, and scattered off of, the reflective metalens. Also shown is the design focal plane in which the intensity was measured. C- Intensity, normalized by the maximum, in the focal plane (z = -11μm) of a conformally layered multispectral metalens illuminated at normal incidence. Overlaid reflected focusing efficiency (not equal to the complete focusing efficiency discussed earlier) is measured by the power within the diffraction-limited spot (circled in red) relative to the reflected power within the 20 μm by 20 μm range pictured (measured at a distance of 1 μm from the lens). The lens was optimized for the target wavelengths pictured (chosen to integrate with existing multispectral filter array designs[17]), and was a square of size 2 μm by 2 μm. Note the decreased performance with decreasing wavelength – likely due to the increased horizontal mixing of lower wavelengths between adjacent meta-atoms.

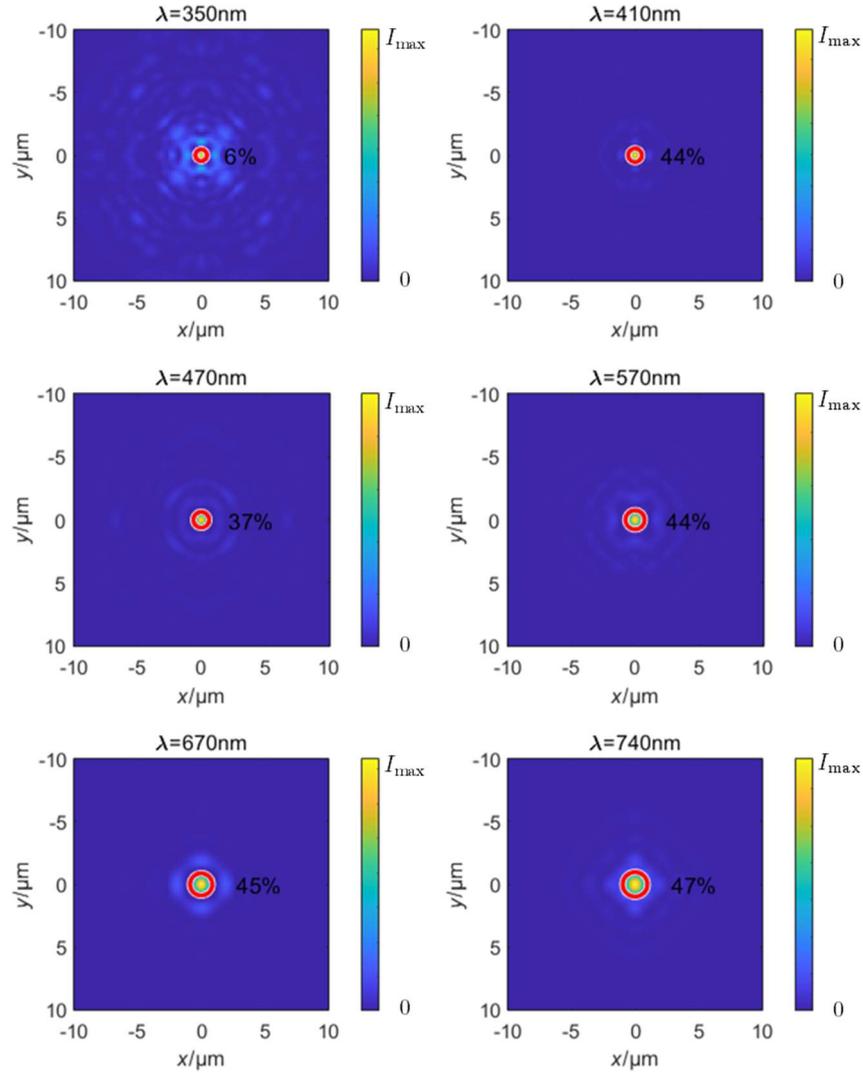

Figure 6. The form of the figures is identical to the previous figure (5.C), but optimized for slightly different wavelengths, and for a larger lens, of size 3 μm by 3 μm. Note that, although only a single wavelength was changed, and only by 10 nm in value, the distribution of the focusing efficiency between the optimized wavelengths was significantly altered.

Although the figures show significant focusing efficiency (an average of 48% and 37% for each lens respectively, across 350-740 nm) this was not relative to the total power incident on the metalens, and so should be expected to be an overestimate. Further, it can be noted that as the wavelength of light is reduced, the focusing efficiency of the realized lens is typically decreased. We suspect this is due to an increase in scattering from the sharp discontinuities between meta-atoms, resulting in increased coupling, something which could be reduced if the forward design process accounted for the thin-film cavities with some gradient (proposed in 3.3). Nonetheless, the key result of these figures is the demonstration that, despite the myriad approximations used to reach this point—including choosing the set of meta-atoms least well-suited to the forward design process used—multispectral focusing with non-negligible efficiencies over relatively large numerical apertures is still achieved and acts to provide an alternative design scheme for conventional nanoscale meta-atom based metalenses.

## 3.3 Use of conformal multilayer structures for metalenses

The appearance of roughly continuous layers was one of the initial assumptions made in designing this metalens, and so to see it in the final structure is relatively unsurprising. In fact, this might even be expected whenever the properties of a material are desired to vary continuously. The particular form of these layers is, however, interesting, and certain features are highlighted in Figure 7A. The fact that, in the limit of large OPDs, a somewhat regular pattern appears to emerge is intriguing, especially since the complete structure should be expected to be aperiodic due to incommensurate length scales of the wavelengths involved. Unfortunately, it is for these increased OPDs that the rotation the wavefront is required to undergo is so extreme that that approximation of no horizontal wavefront motion is necessarily inapplicable, perhaps suggesting that one-dimensional delay lines with constant curvature, rather than simply vertical delay lines, should be used in future work. A brief glance at the apparent 'noise' in the structure of this lens also suggests another drawback to this analysis, which is the sensitivity of the minimum found by the optimization routine to its initial conditions. We further note that the emergence of common underlying features can be exploited for manufacturing ease and used as an initial starting point for future work.

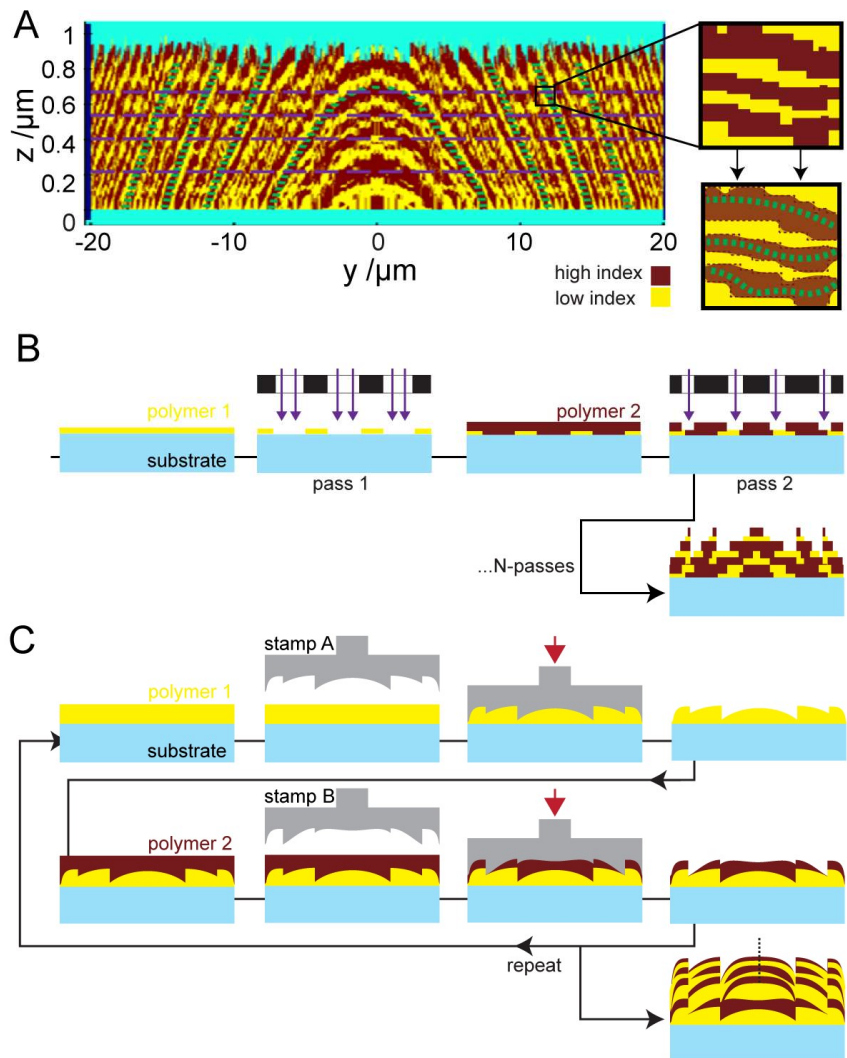

Figure 7: A- Figure 4A reproduced, but with additional lines marked to highlight particular structures. The green dotted line highlights the surface of layers of $TiO_2$ which appear qualitatively similar in form to a hyperbolic graph (albeit modulo

1 μm) such as the path difference profile under consideration, whilst the purple dashed lines trace a slightly weaker pattern – that of layers where $TiO_2$ appears more often than $MgF_2$. Both of these features can be exploited for manufacturing ease and as a starting point for future work. B- Example of manufacturing using multi-pass exposures similar to multi-level diffractive optical element (DOE) production and capable of producing sharper features. C- Example of cost-effective manufacturing approach of conformal multilayer metalens with a step-and-repeat molding process using high-low index polymer thin-films. We note that multiple stamps would likely be required.

The benefits of conformally layered structures are significant. Conventional miniaturized lenses are typically manufactured through diamond turning, molding or multi-step optical lithography (e.g. multi-level phase diffractive optics), and a jump to high-cost ultra-high resolution patterning using electron beam lithography (or otherwise) to manufacture highly sub-wavelength meta-atoms (with tight tolerances) for metalenses that provide similar performance to state-of-the-art cost-effective refractive lenses is somewhat of a non-starter. Alternatively, as multi-layer thin film optical coating and molding / imprinting micro-optics are both relatively mature technologies/ industries, one can envisage the manufacture of multi-layer metalenses with alternating high-low index polymers (potentially with metallic/dielectric inclusions) deposited / spin-coated, and molded to shape, likely a step-and-repeat process. Figure 7B and 7C show two example manufacturing approaches, the first (B) is a multi-pass exposure approach akin to the production of multi-level diffractive optics using N-binary masks (or $2^N$ depending on mask design / exposure pattern). One can envisage a photomask with each sub-region dedicated the $N^{th}$ pattern and the polymers different photoresists (or photoresists with embedded dielectric inclusions). The second approach, (C), for manufacturing is to use multiple master stamps (molds) and an embossing process with alternating polymer layers—a process similar to the cost-effective replication of plastic micro-optic elements. We note that this is not exhaustive, and there are many potential means to manufacture multi-layer metalenses including vacuum polymer deposition over a master structure (akin to conformal parylene coatings), a modification on grayscale lithography with multiple thin-film layers, and atomic layer deposition.

In comparison to sharp discretization, the notion of smooth layers increases the ease of manufacturing, as the number of sharp features required in a conformal design is greatly reduced. Although the natural waveguide dispersion of discrete meta-atoms is lost in such an example, this dispersion typically acts counter to the properties desired for a lens, rather than complementarily to it. The meta-atoms also become conceptually simpler, and it is likely that approximation schemes could be developed for conformally layered surfaces of low curvature. Further, an analogy can be drawn to the multi-layer coatings atop of conventional refractive lenses / optics for spectral filtering functionality (i.e. bandpass, notch and edgepass filters), and our approach may be conceptualized as a further extension to this. Conformally layered structures also provide additional degrees of freedom without the requirement of additional sharp features being added into a meta-atom. The use of the adjoint-state method[6] to optimize surfaces to achieve a particular focusing efficiency also provides a fast route for inverse design, whilst the forward design processes presented here, and improvements on them, would allow for initial conditions for the inverse design process to be rapidly calculated.

One fundamental disadvantage is present, however: coupling along the thin-films is almost guaranteed, hugely increasing the difficulty of forward design without the use of the drastic approximation regimes such as those presented here and metasurfaces with discontinuous properties also become particularly difficult to realize, due to the same coupling. Additional disadvantages include the fact that, due to overlaps, although the number of layers may be bounded at any point, an arbitrary number of continuous layers may be required to completely cover a lens, which may in fact necessitate some discontinuities during manufacturing. Lastly, it is noted by the authors that a motivation of single-layer metalenses in the literature is to replace refractive or complex shaped optics with minimal surface engineering, hence we acknowledge that 'building back up' once more may appear counter to this thrust in the community. We would argue that we are compromising between conceptually attractive but difficult to manufacture metalenses (nanoscale meta-atom designs) and bulky refractives with excellent optical performance but produced in cost-effective ways (including applied thin-film coatings), whilst the resulting lens remains ultrathin.

## 4. CONCLUSIONS

This work considered the bounds, some presumed, and some proven, on the achromaticity of a metalens for any given refractive index and thickness, and demonstrated how they can be avoided, with the underlying limit being related to the spectral control which can be induced and the trade-off in intensity from all the areas where such control cannot be achieved. It was noted how a more general optimization routine would allow this additional freedom to be used in the forward design of metalenses with arbitrary meta-atom libraries, as well as the limitations of this approach. It was then noted that, as the thin-film meta-atom library selected lacked the sufficient spectral control to make true achromaticity viable, in most imaging applications multispectral imaging is sufficient. Thus, a metalens was designed based on thin-film dielectric stacks, produced using the optimization routine described, resulting in high efficiency diffraction-limited focusing of the reflected light, on the order of >40% on average across all wavelengths (350-740 nm). It is demonstrated that this optimization routine gives rise to a series of conformal layers, with relatively large-scale properties.

The consequences of a general design based around conformal layers is considered, with particular advantages in manufacturing and design (such as allowing rapid first approximations to a more complete inverse-design problem), and additional advantages in more general forward design algorithms, if local coupling between meta-atoms can be accounted for. Potential disadvantages arise in the production of very large-scale structures, however, as well as in producing an entirely separable forward design process, something which becomes difficult. Some promising directions for future research identified include:

- The form of a multilayered lens in the limit of large OPDs.
- Finding a family of 1D paths to retrieve approximate separability and forward design (perhaps following the arc of a circle, due to its constant curvature, rather than the constant normal to the metasurface assumed here).
- Considering the local effects of a finite gradient on thin-film metasurfaces.
- Applying adjoint-state inverse design optimization, with forward design to provide initial conditions.
- Applying calculation of focusing efficiency project directly from a given transmission profile to the farfield to avoid reliance on matching to an existing profile.

## ACKNOWLEDGEMENTS

The authors would like to acknowledge the support of: the University of Cambridge Undergraduate Research Opportunities Program (UROP); the Wellcome Trust (Junior Interdisciplinary Fellowship); the Engineering and Physical Sciences Research Council (EP/R003599/1); and Cancer Research UK (A24669, C55962).## REFERENCES